\documentclass[prl,twocolumn,aps,byrevtex,showpacs,tightenlines]{revtex4}

\begin{document}

\title{Quantum coherence in the presence of unobservable quantities}

\author{Kae Nemoto$^{1,2}$ and Samuel L. Braunstein$^2$}
\affiliation{${}^1$National Institute of Informatics, Tokyo 101-8430,
Japan\\}
\affiliation{${}^2$Informatics, Bangor University, Bangor LL57 1UT, UK}

\date{\today}

\begin{abstract}
State representations summarize our knowledge about a system. When 
unobservable quantities are introduced the state representation
is typically no longer unique. However, this non-uniqueness does not affect
subsequent inferences based on any observable data. We demonstrate that 
the inference-free subspace may be extracted whenever the quantity's 
unobservability is guaranteed by a global conservation law. This 
result can generalize even without such a guarantee. In particular,
we examine the coherent-state representation of a laser where the 
absolute phase of the electromagnetic field is believed to be unobservable.
We show that experimental coherent states may be separated from the 
inference-free subspaces induced by this unobservable phase. These 
physical states may then be approximated by coherent states in a 
relative-phase Hilbert space.
\end{abstract}


\maketitle

\section{introduction}

The representation of a state and its associated interpretation 
are fundamental issues in quantum mechanics. The state representation
of a system summarizes our knowledge about that system; it summarizes the
information about any observable we wish to measure.  Conversely, 
experiments on a system allow us to determine the associated state 
representation. 

Now it turns out that some observables are not accessible to
experiment. Such quantities might be called {\it unobservables}, a
term which we shall take to be a synonym for unmeasurable quantities.
There are several mechanisms by which a quantity may be unobservable.
One rigorous mechanism involves global conservation laws. In
particular, the Wigner-Araki-Yanase (WAY) theorem says that any
operator which does not commute with an operator corresponding to a
global conservation law is not observable \cite{Wigner52, Araki60}. 
Therefore any system satisfying global conservation laws involves
unobservable quantities.

In addition to those quantities whose unobservability is directly
associated with global conservation laws, there are other quantities
for which no rigorous argument currently exists guaranteeing their
unobservability, but which are, nonetheless, generally accepted to be
inaccessible to any experimental means. An example of such a quantity
is the absolute phase of the electromagnetic field \cite{Moelmer97}.

What are the consequences for the state representation of a system when
unobservable quantities are involved? Since such a situation implies
that there will be parts of the state's representation which cannot be
examined this means that the state could equally well be represented
in various functionally indistinguishable though distinct forms. Thus,
an apparent consequence of dealing with unobservables is that the state
representation is effectively non-unique.

In this paper we investigate the consequences of unobservable 
quantities on the state representation more fully. To do this, we 
consider the possibility that the non-uniqueness can be isolated
in the state representation.  More particularly, that 
the state may be written in a form of a direct product of two states
or components, where at least one of these components involves no 
non-uniqueness due to unobservability. This component involves a
subspace which is spanned by only eigenstates of truly observable
quantities. By contrast, the minimal non-unique component corresponds
to that part of the state representation which cannot be determined by
any experimental procedure; it corresponds to an inference-free
subspace.
 
In section 2 we start by considering the case where the
unobservability of quantities is guaranteed by global conservation
laws and demonstrate that the inference-free subspace may 
be isolated and separated from the remainder of the state
representation. A hint of how this may be achieved can be found in the
WAY theorem itself which also suggests that the unobservability of the
non-commutative operators can be bypassed by taking into account their
own relative quantities. We use the WAY theorem together with a
consideration of state preparation to show that a relative-quantity
Hilbert subspace can be constructed. In section 3, we consider a
generalization of this argument, which may be applied to other cases
involving unobservable quantities.  In particular, we look into the
case of the laser output field, where its absolute phase is generally
accepted to be unobservable, though a direct proof for this based on
the WAY theorem is currently lacking. Notwithstanding this, we demonstrate
how the inference-free component of the state description may be
isolated as an excellent approximation.

\section{the Wigner-Araki-Yanase theorem and an inference-free
subspace}

The Wigner-Araki-Yanase (WAY) theorem gives us a playground of
systems where unobservability of certain quantities is guaranteed by
global conservation laws. The WAY theorem states that any operator
which does not commute with an operator of the global conservation is
not observable \cite{Wigner52, Araki60}. Consider a system which
consists of the observed subsystem and its measuring apparatus. As the
total momentum $\hat{\Pi}$ of the system is conserved, a position
operator $\hat{x}$ of the observed subsystem is unobservable. This is
because the position operator does not commute with the total momentum
and such measurement process violates the conservation law. 

Now we give a way to construct a relative-quantity subspace.  
According to the WAY theorem, a relative quantity of the unobservable
absolute operators can be observable.  By constructing a subspace
where the relative quantity can be well-defined, we isolate the
inference-free component.
Here we show an example which can be easily generalized.
A relative position operator of the observed system to the apparatus
$\hat{x}_1-\hat{x}_2$ ($=\hat{x}_r$) commutes with the total momentum
and hence is observable, where $\hat{x}_1$ and $\hat{x}_2$ are the
absolute positions of the observed system and of the apparatus respectively. 
We take eigenstates of an operator $\hat{x}_a=\hat{x}_1+\hat{x}_2$ to
construct the entire Hilbert space together with eigenstates of
$\hat{x}_r$. The Hilbert space for the entire system (the observed
system and the apparatus) can be expanded by
$\{|x_r\rangle\otimes|x_a\rangle\}$ as well as by 
$\{|x_1\rangle\otimes|x_2\rangle\}$. To construct a relative-position
Hilbert space, we start with separable states given by 
\begin{eqnarray} \label{separable}
|\psi\rangle = |\psi_r\rangle \otimes |\psi_a\rangle \;,
\end{eqnarray}
where
\begin{eqnarray}
\begin{array}{l}
\left\{ \begin{array}{l}
		 |\psi_r\rangle=\int dx_r  \psi_r(x_r)|x_r\rangle\\
		|\psi_a\rangle=\int dx_a \psi_a(x_a)|x_a\rangle.\!
	\end{array}
\right.
\end{array}
\end{eqnarray}
Here the state is separable in terms of the two subspaces of
$\{|x_r\rangle\}$ and $\{|x_a\rangle\}$.

As the operator $\hat{x}_a$ is not observable, the state $|x_a\rangle$
must be considered as a label of the equivalence class of states
\cite{KS}, hence the state $|x_a\rangle\langle x_a|$ implies a set of
the states  
\begin{equation}
\int dXP(X)e^{-iX\Phi}|x_a\rangle\langle x_a|e^{iX\Phi},
\end{equation}
with all possible prior distributions $P(X)$.
Using this representation, the total state can be represented as 
\begin{equation} \label{XX}
\rho_{ra} = 
\int dX P(X) e^{-iX\hat{\Pi}} |\psi\rangle\langle \psi| e^{iX\hat{\Pi}}.
\end{equation}
The operator $\hat{x}_r$ commutes with the total momentum
$\Pi$, then the state $|\psi_r\rangle$ is preserved under the
action of the displacement operator $e^{-iX\Phi}$. This allows the
density matrix to be  
\begin{eqnarray} \label{sep}
\rho = |\psi_r\rangle\langle \psi_r| \otimes \rho_a.
\end{eqnarray}
where
\begin{eqnarray}
|\psi_r\rangle &=& \int dx_r \psi_r(x_r)|x_r\rangle\nonumber\\
\rho_a &=& \int\int\int dX P(X) dx_adx'_a
\psi_a(x_a)\psi^*_a(x'_a)\nonumber\\
&& \times e^{-iX\hat{\Pi}}|x_a\rangle\langle x'_a|
e^{iX\hat{\Pi}}.\nonumber
\end{eqnarray}
The relative-position state $|\psi_r\rangle$ is on the
relative-position Hilbert space and the relative-quantity operators
can be defined on this subspace. The state $\rho_a$ constructs an
inference-free component and the inference-free subspace is
constructed to be completely free from a choice of the prior
distribution. 

Now, let us generalize the argument to entangled
states. The general state can be represented by
\begin{equation}\label{entangled}
|\phi\rangle = \int dx_r dx_a \psi(x_r,x_a) |x_r, x_a\rangle.
\end{equation}
In the case where the state is entangled, the function $\psi(x_r,x_a)$
cannot be written as $\psi_r(x_r)\psi_a(x_a)$. This state in the same
representation with a prior is  
\begin{eqnarray} \label{ent}
\rho &=& \int\cdots\int dx_a dx_{a'} dx_r dx_{r'} \nonumber \\
&&\phantom{ \int\cdots\int}\times
\psi(x_r,x_a)\psi^*(x_r',x_a')|x_r\rangle \langle x_r'|\nonumber\\ 
&&\phantom{ \int\cdots\int}
\otimes\! \int dX P(X) |x_a+X\rangle \langle x_a'+X|\;.
\end{eqnarray}
By contrast to the separable case, it seems non-trivial to construct
an inference-free subspace.  
Such entangled states can be obtained by assuming arbitrary separable
states and some entangling operators.  For instance, a product state
of $|x_r\rangle$ and a superposition of the total momentum eigenstates
is separable by definition and yet can generate entanglement  
with some entangling operator such as a SUM gate 
($\exp \big(-i \hat{x}_r \otimes \hat{\Pi}\big)$).  In fact, the SUM
gate commutes with the total momentum and hence such an operation is
allowed, so it seems that we can create an entangled state not
violating the conservation low.  However, the essential issue in here
is to consider the state representation process to obtain a consistent
state representation under the global symmetries.  Next we will show that 
the global symmetries impose restrictions in the state representation
process.

As we have discussed above, a superposition can generate entanglement
with some entangling operator, while by any of the allowed operations
an eigenstate of the total momentum cannot be entangled with the
relative-position subspace. This leads us to a question if any creation of
supposition can be allowed under the global symmetries.
It is not difficult to see that none of the operators which
generate a superposition from an eigenstate is allowed under the
conservation law.  Any creation of superposition necessitates 
a third system to be involved in the state preparation process. 
This is inconsistent with the global symmetries. 
This concludes that considering the state preparation process, only
the eigenstates of the total momentum are consistent with the global
symmetry.  This constraints on states allowed in the system could be
considered as a superselection rule. The original work by Wigner
\cite{Wigner52} and following works \cite{ozawa} have allowed the
system to prepare an arbitrary state, in particular superpositions so
that measurement of non-observables in the sense of the WAY theorem
can be arbitrarily precisely done.  However, even if the
system of the observed system and the apparatus recovers the
conservation of the total momentum after the state preparation, the 
system cannot completely eliminate the third system.  
For example, a closed system with the momentum conservation is invariant in
transformation by its absolute position, so different values of the
total momentum gives the same state to the system.  Two different
values of the {\it total momentum} $\hat{\Pi}$ become distinct when these
are realized in the extended system.
Thus, the extended system is necessary for the physical meaning of
superpositions and the superposition states have to be captured in a
relative-quantity subspace in the extended system.
For a closed system with the momentum conservation, as the eigenstate
of the total momentum is the only state consistent, any state can be
represented as (\ref{separable}) and hence the relative-position
subspace always can be constructed.

\section{The case of laser output field}
In this section, we generalize the argument to cases where
unobservability is not guaranteed by global conservation laws. Our
particular interest of such cases here is laser output field.
Despite a lack of rigorous proof, the absolute phase of an 
electromagnetic field has been considered to be non-observable
\cite{Moelmer97,Gea}.  
Due to the non-observability of absolute phase $\phi$, 
the state representation for the laser output field has an inference
in terms of this quantity, and is written as  
\begin{eqnarray} 
\label{phi}
\rho = \int_0^{2\pi} \frac{d\phi}{2\pi} P(\phi) \big|\, |\alpha|
e^{-i\phi}\big\rangle\big\langle |\alpha| e^{-i\phi} \big|
\label{coherent_s} \\
\end{eqnarray}
where $P(\phi)$ is an untestable prior distribution function, which is
inference in the state representation.
Now we take the state representation of laser output field as an
example of the general case to consider construction of the
relative-quantity subspace. The previous argument suggests to generalize
the state to two mode and take a relative phase of a two-mode coherent
state.  However the argument about state preparation does not apply
here, as unobservability of absolute phase being a weaker condition
than global conservation law, then we have to find an alternative way
to construct an inference-free subspace.

A two-mode coherent state is given as 
\begin{eqnarray} \label{ab}
|\alpha, \beta\rangle &=& \big|\; |\alpha| e^{-i\phi_{\alpha}}\rangle \otimes 
\big|\; |\beta|e^{-i\phi_{\beta}}\rangle \nonumber\\
&=& e^{-\frac{|\alpha|^2+|\beta|^2}{2}} \sum_{n_1}^{\infty}\sum_{n_2}^{\infty}
\frac{\alpha ^{n_1}\beta^{n_2}}{\sqrt{n_1 ! n_2!}}|n_1,n_2\rangle.
\end{eqnarray}
The total photon number of the state is $N=n_1+n_2$ 
and the difference photon number is $M=(n_1-n_2)/2$
which is either integer (for even total photon numbers) or
half-integer (for odd total photon numbers).  The state (\ref{ab}) can
be alternatively expanded by the eigenstates characterized by these
quantum numbers $N$ and $M$ as
\begin{eqnarray} \label{twomode-nm}
|\alpha,\beta\rangle &=& e^{-\frac{\langle\hat{N}\rangle}{2}} 
\sum_{N=0}^{\infty} \sum_{M=-\frac{N}{2}}^{\frac{N}{2}} 
\frac{\alpha^{\frac{N}{2}+M}
\beta^{\frac{N}{2}-M}}
{\sqrt{\big(\frac{N}{2}+M\big)! \big(\frac{N}{2}-M\big)!}} \nonumber \\
&&\phantom{e^{-\frac{\langle\hat{N}\rangle}{2}}
\sum_{N=0}^{\infty} \sum_{M=-\frac{N}{2}}^{\frac{N}{2}}}\times|N,M\rangle\;.
\end{eqnarray}
where $\langle\hat{N}\rangle=|\alpha|^2+|\beta|^2$.
Obviously this state is not separable in terms of the two subspaces,
$\{|N\rangle\}$ and $\{|M\rangle\}$. Unobservability of absolute phase
is weaker as a restriction to the system, so the superrule for the
conserved system does not apply here.  Such a case, in general, does
not allow us to simply isolate an inference-free subspace and requires
an ingredient to approximately do so. In this case of coherent states,
we take large total photon number limits to construct a relative-phase
subspace.  

Taking a set of parameters as 
\begin{eqnarray}\label{para}
\frac{|\alpha|}{\langle\hat{N}\rangle^{1/2}}&=&-\sin\frac{\theta}{2}\nonumber\\
\frac{|\beta|}{\langle\hat{N}\rangle^{1/2}}&=&\cos\frac{\theta}{2}\nonumber\\
	\phi_{\alpha}-\phi_{\beta}&=&\phi_r\;.
\end{eqnarray}
The two mode coherent state can be written as the sum of spin coherent
states, yielding
\begin{eqnarray} \label{twomode}
|\alpha,\beta\rangle = e^{-\frac{\langle\hat{N}\rangle}{2}} 
\sum_{N=0}^{\infty}
\frac{\big(\langle\hat{N}\rangle^{1/2}e^{-i\phi_{\beta}}\big)^N}{\sqrt{N!}}
|N\rangle\otimes |\theta, \phi_r\rangle_N.
\end{eqnarray}
Here $|\theta, \phi_r\rangle_N$ is a spin-$N/2$ coherent state with
the parameterization (\ref{para}). 
Alternatively the spin coherent state may be parameterized by $\xi$ ($= -
\frac{|\alpha|}{|\beta|}e^{-i\phi_r}$) as
\begin{eqnarray}
|\theta, \phi_r\rangle_N &=&|\xi\rangle_N \nonumber\\
&=& \sum_{M=-\frac{N}{2}}^{\frac{N}{2}}
\left( 
	\!\!\begin{array}{c}
	N\\
	\frac{N}{2}-M
	\end{array}\!\! \right)^{\!\frac{1}{2}}\nonumber\\
&&\phantom{\sum_{M=-\frac{N}{2}}^{\frac{N}{2}}}
\times(1+|\xi|^2)^{-\frac{N}{2}}\xi^{\frac{N}{2}+M} |M\rangle\;.
\end{eqnarray}
If the spin coherent state $|\xi\rangle_N$ is not dependent on the
total photon number $N$, then the state for the total
photon number can be realized as a coherent state of
$|\langle\hat{N}\rangle^{1/2} e^{-i\phi_{\beta}}\rangle$. 

Here we consider a limit of large total photon number, 
$\langle\hat{N}\rangle^{1/2} \to \infty$. The contribution of
components for small $N$ to the sum is negligible and the main
contribution is the terms of the order 
$N \simeq \langle\hat{N}\rangle^{1/2}$. In the large limit of $N$, the spin
coherent state can be contracted to a Weyl-Heisenberg (WH) coherent state.
When $|\alpha|\simeq |\beta|$, the state can be typically 
contracted to a WH coherent state, 
\begin{eqnarray}
|\theta,\phi_r\rangle \to |-\sqrt{2}|\alpha|e^{-i\phi_r}\rangle.
\end{eqnarray}
At the limit, this coherent state is approximately separable with the
subspace of the total photon number, and can be extracted from the sum
in Eq. (\ref{twomode}) as 
\begin{eqnarray}
|\alpha,\beta\rangle &\simeq& |-\sqrt{2}|\alpha|e^{-i\phi_r}\rangle
\nonumber \\
&&\otimes \,
e^{-\frac{\langle\hat{N}\rangle}{2}} 
\sum_{N=0}^{\infty}
\frac{\big(\langle\hat{N}\rangle^{1/2}e^{-i\phi_{\beta}}\big)^N}{\sqrt{N!}}
|N\rangle\nonumber\\
&=& |-\sqrt{2}|\alpha|e^{-i\phi_r}\rangle\otimes 
|\langle\hat{N}\rangle^{1/2} e^{-i\phi_{\beta}}\rangle\;.
\end{eqnarray}
The laser output state $\rho$ in the equivalent class of this state with a
prior $P(\phi_\beta)$ may be 
given as
\begin{eqnarray}
\rho &=& \int d\phi_\beta P(\phi_\beta) \Big( 
|-\sqrt{2}|\alpha|e^{-i\phi_r}\rangle\langle-\sqrt{2}|\alpha|e^{-i\phi_r}|
\nonumber\\
&&\otimes
|\hat{N}\rangle^{1/2} e^{-i\phi_{\beta}}\rangle
\langle\langle\hat{N}\rangle^{1/2} e^{-i\phi_{\beta}}|\Big) \;.
\end{eqnarray}
The space of the relative phase $\phi_r$ is independent from the
integral of the absolute phase $\phi_\beta$, yielding
\begin{eqnarray} 
\rho =|-\sqrt{2}|\alpha|e^{-i\phi_r}\rangle\langle-\sqrt{2}
|\alpha|e^{-i\phi_r}| \otimes \rho_{\strut \rm P}\;,
\end{eqnarray}
where 
\begin{eqnarray}
\rho_{\strut \rm P} = \int d\phi_\beta P(\phi_\beta) 
|\hat{N}\rangle^{1/2} e^{-i\phi_{\beta}}\rangle
\langle\langle\hat{N}\rangle^{1/2} e^{-i\phi_{\beta}}|\;.
\end{eqnarray}
Hence the relative-phase subspace can be approximately
constructed.  However this
approximation is not so useful as the limit brings $|\alpha|$ also to
infinity as $\sqrt{2}\alpha \simeq \langle\hat{N}\rangle^{1/2}$.  
By contrast, when $|\alpha|<<|\beta|$ is satisfied, the
group contraction may be taken in the order of $\langle
\hat{N}\rangle$. In this case the spin state $|\xi\rangle_N$ is
contracted by a parameter $\epsilon = 1/|\beta|$ as
\begin{equation}
\xi=-\epsilon |\alpha|e^{-i\phi_r}, \; (\epsilon \to \infty)\;.
\end{equation}
In this contraction, the spin size given by $|\beta|^2$ goes to
infinity with $\epsilon \to 0$ and the state is contracted to a
WH coherent state $|-|\alpha|e^{-i\phi_r}\rangle$.  

The coherent state from laser can be approximately represent as
\begin{equation} \label{relative}
|\alpha,\beta\rangle\langle\alpha,\beta| \simeq
|-|\alpha|e^{-i\phi_r}\rangle\langle -
|\alpha|e^{-i\phi_r}|\otimes \rho_{\strut \rm P}\;,
\end{equation}
under the condition 
\begin{equation}\label{cond}
\langle\hat{N}\rangle \simeq |\beta|^2 \gg |\alpha|^2.
\end{equation}
Hence the state on the subspace of the relative phase is a coherent
state, which is, in fact, what we call a coherent state in experiments.

To conclude, we have shown the explicit construction of an approximate
relative-phase Hilbert space. The two mode coherent state can be
represented as a pure coherent state in the relative-phase subspace
under the condition (\ref{cond}).  This state presentation of relative
phase does not involve prior distribution, and hence
circumvents the entire discussion about unknowable absolute phase.

\vskip 0.1truein
KN acknowledges financial support form the Sumitomo Foundation and the
hospitality of Hewlett Packard Labs in Bristol, UK. 
SLB currently holds a Wolfson-Royal Society Research Merit Award.

\end{document}